\documentclass[aps,prl,groupedaddress]{revtex4}

\usepackage{graphicx}

\begin{document}

\title{Signature of the electron-electron interaction in the magnetic field
dependence of nonlinear I-V characteristics in mesoscopic conductors}
\author{E. Deyo}
\affiliation{Physics Department, University of Washington, Seattle, WA 98195, USA}
\author{A. Zyuzin}
\affiliation{A.F.Ioffe Institute, 194021 St.Petersburg, Russia}
\author{B. Spivak}
\affiliation{Physics Department, University of Washington, Seattle, WA 98195, USA}

\begin{abstract}
The nonlinear I-V characteristics of mesoscopic samples contain parts which
are linear in the magnetic field and quadratic in the electric field. These
contributions to the current are entirely due to the electron-electron
interaction and consequently they are proportional to the electron-electron
interaction constant. We present detailed calculations of the magnitude of
the effect as a function of the temperature, and the direction of the magnetic
field. We show that in the case of a magnetic field oriented parallel to the
sample, the effect exists entirely due to spin-orbit scattering. The
temperature dependence of the magnitude of the effect has an oscillating
character with a characteristic period on the order of the temperature itself.
We also clarify in this article the nature of the electron-electron
interaction constant which determines the magnitude of the effect.
\end{abstract}

\maketitle

\section{Introduction}

According to Onsager, the linear conductance $G(\mathbf{H})$ of a conductor
measured by the two-probe method must be an even function of the magnetic
field $\mathbf{H}$ \cite{Landau}: 
\begin{equation}
G(\mathbf{H})=G(-\mathbf{H})  \label{Landau}
\end{equation}%
This is a consequence of general principles: the time reversal symmetry and
the positive sign of the entropy production. Therefore it holds in all
nonmagnetic conductors. In a single particle approximation and at zero
temperature the validity of Eq. \ref{Landau} also can be verified using the
Landauer formula for the conductance of a sample 
\begin{equation}
G=\frac{e^{2}}{h}\sum_{ij}|T_{ij}(\mathbf{H})|^{2}  \label{landauer}
\end{equation}%
and requirement of time reversal symmetry $T_{ij}(\mathbf{H})=T_{ji}^{\ast
}(-\mathbf{H)}$.  Here $T_{ij}$ is a scattering matrix between
electronic channels labelled by the indices $i$ and $j$ (see for review of
the subject Ref. \cite{binnakker}).  To verify the validity of Eq. \ref{Landau}
in mesoscopic samples \cite{AltshulerKhmelnitski} one has to prove that $
\langle (G(\mathbf{H})-G(-\mathbf{H}))^{2}\rangle =0$, which requires the
cancellation of the odd-in-$\mathbf{H}$ parts of the Cooperon and Diffuson type of
diagrams shown in Fig. 1h,i,j. Here the brackets $\langle \rangle $ denote
averaging over random realizations of the impurity potential (we use
standard diagram technique for averaging over random scattering
potential \cite{Abricosov}). 

On the other hand there are no general
principles preventing the existence of odd-in-$\mathbf{H}$ terms in the
nonlinear I-V characteristics of conductors. In this article we study the
quadratic-in-voltage $V$ part of the I-V characteristics which can be
represented as 
\begin{equation}
I_{(nl)}=V^{2}[F_{o}(\mathbf{H})+F_{e}(\mathbf{H})]  \label{eq:InlGen}
\end{equation}%
where $F_{o}(\mathbf{H})$ and $F_{e}(\mathbf{H})$ are odd and even functions
of $\mathbf{H}$ respectively. Since $\mathbf{H}$ is an axial vector and the
current density, $\mathbf{j}$, is a polar one, the function $F_{o}(\mathbf{H})$ can be non-zero
only in non-centro-symmetric media. It is important to study $F_{o}(\mathbf{H%
})$ because of the fact that in the approximation of noninteracting
electrons, $F_{o}(\mathbf{H})=0$. It is particularly simple to verify this
fact using the Landauer formula. Indeed, in the absence of the
electron-electron or electron-phonon interactions, the total current through
the sample will be the sum of contributions from different electron
energies. Each of these contributions is an even function of $\mathbf{H}$,
and hence the total current will also be an even function
of $\mathbf{H}$. Thus the effect is entirely due to electron-electron or
electron-phonon interaction. In contrast, the even-in-$\mathbf{H}$ function $%
F_{e}(\mathbf{H})$ can be described even in a single particle approximation
(see, for example, Ref. \cite{LarkinKhmelnitskii} where the calculations were
done in the case of mesoscopic samples).

In the case of pure bulk non-centro-symmetric crystals, and at high
temperatures the effect described by $F_{o}(\mathbf{H})$ has been
investigated both theoretically and experimentally (see, for example, Ref. 
\cite{Photovoltaic}). In the case of chiral carbon nanotubes a classical
theory of this effect was discussed in Ref. \cite{ivchenko}. At high
temperatures there are two contributions to the effect:

a) The first contribution is purely classical and it can be described in the
framework of the Boltzmann kinetic equation: An electric field accelerates
electrons creating a non-equilibrium distribution function. This non-equilibrium distribution function consists of two parts: 
an anisotropic-in-momentum part that is proportional to $V$ and a quadratic-in-$V$ contribution which is isotropic in the
momentum. In isotropic media, the relaxation of this non-equilibrium
distribution due to inelastic scattering processes yields no net current;
however, in non-centrosymetric media and in the presence of the magnetic
field, the inelastic relaxation rate has odd in electron momentum components,
which give rise to the odd-in-$\mathbf{H}$ part of Eq. \ref{eq:InlGen} described
by $F_{o}(\mathbf{H})$.

b) The second contribution \cite{BelinicherIvchenko} is due to a shift in
the center of mass of a wave-packet during collisions. The description of
these processes is beyond the classical Boltzmann kinetic equation. In
non-equilibrium and non-centro-symmetric media and in the presence of the
magnetic field these shifts take place in a particular direction determined
by a crystal symmetry, and lead to an odd in $\mathbf{H}$ contribution in
the net current through the sample. This contribution is similar to that
discussed in the framework of the anomalous Hall effect \cite{Hall}. The
above two contributions to $F_{o}(\mathbf{H})$ are proportional to the
inelastic electron relaxation rate $1/\tau_{\epsilon}$, and consequently,
they vanish at $T=0$ (there are of course contributions from the above
effect to the I-V characteristics which at $T=0$ are proportional to higher
powers of $V$).

It has been pointed out in Refs. \cite{Buttiker,SpivakZyuzin1} that in
mesoscopic metallic samples, where all spatial symmetries are broken, there
is an odd-in-$\mathbf{H}$ contribution to Eq. \ref{eq:InlGen} which survives
in the limit $T=0$ and which therefore determines the magnitude of the
effect at small $T$. This effect has been observed experimentally \cite
{Cobden,Markus,Leturcq,Marlow,Helene}. As usual for mesoscopic effects, this
contribution is due to random electronic interference. Therefore it exhibits
random sample specific oscillations as a function of the external magnetic
field, temperature and the electron chemical potential. The characteristic
feature of the effect is that it is proportional to the amplitude of the
electron-electron interaction rather than the scattering rate. The
qualitative explanation of the effect is the following \cite{SpivakZyuzin1}:
The linear in $V$ mesoscopic fluctuations of the current density are due to
random interference of electron waves travelling along different diffusive
paths. Though the total current through the sample should be an even
function of $\mathbf{H}$, the local current densities contain a part which
is odd-in-$\mathbf{H}$.  By the same token, there is a part of the electron
density $\Delta n(\mathbf{r},V, \mathbf{H})$ which is proportional to $V$
and odd-in-$\mathbf{H}$ \cite{SpivakZyuzin}.  In the presence of the
spin-orbit scattering the applied voltage also induces local fluctuations of
the spin density $\Delta \mathbf{S }(\mathbf{r},V,\mathbf{H})$.  These nonequilibrium densities create an additional random potential due to electron-electron interaction 
\begin{equation}
\Delta u(\mathbf{r},V,\mathbf{H})= \beta _{eff}^{(1)}\Delta n(\mathbf{r},V,%
\mathbf{H})  \label{DeltaU}
\end{equation}
and an additional exchange magnetic field 
\begin{equation}
\mathbf{h}(\mathbf{r},V,\mathbf{H})= \beta _{eff}^{(2)}\Delta \mathbf{S}
\label{DeltaH}
\end{equation}
were $\beta _{eff}^{(1,2)}$ are interaction constants. We can then calculate
a change of a linear conductance of a sample, $\Delta G=G(\{u(\mathbf{r)}
+\Delta u(\mathbf{r}),\mathbf{h}(\mathbf{r})\})-G(\{u\})$, induced by a
change of the scattering potential given by Eq. \ref{DeltaU}, which gives us 
\begin{equation}
I_{(nl)}=\Delta G[V,\mathbf{H},\{\Delta u(\mathbf{r},V,\mathbf{H}),\mathbf{\
h}(\mathbf{r},V,\mathbf{H})\}]V  \label{Inl}
\end{equation}
Expanding Eq. \ref{Inl} with respect to $\Delta u(\mathbf{r})$ , and $%
\mathbf{h}(\mathbf{r})$ we arrive at Eq. \ref{eq:InlGen}.

In this paper, we present calculations of the magnitude of the effect as
a function of the temperature and the magnetic field. We show that in the case
of parallel magnetic field, the effect exists entirely due to spin-orbit
scattering. The temperature dependence of the magnitude of the effect (both
in parallel and in the perpendicular magnetic field) randomly oscillates with a characteristic period on the order of the temperature itself.
We also take into account the exchange contribution to the effect and
clarify the nature of the electron-electron interaction constant which
determines the value of $F_{o}(\mathbf{H})$ in Eq.\ref{eq:InlGen}.

\section{Diagrammatic calculations of the magnitude of the effect}

Let us consider a sample shown in the insert of Fig.1 with the
characteristic size $L$ which is much larger than the electron elastic mean
free path $l$. In this limit $\langle I_{nl} \rangle =0 $. Therefore we
shall characterize the magnitude of the current by the variance $\langle
I_{nl}^{2}\rangle$. Before averaging over random potential configurations,
to first order in the electron-electron interaction constant, the value of $%
I_{nl}$ is given by the diagram shown in Fig. 1a. After averaging, the
quantity $\langle I_{nl}^{2}\rangle $ is given by diagrams shown in Fig. 1d.
In these diagrams, the solid lines which correspond to the electron Green
functions carry frequencies on the order of the temperature $T$, so we can neglect
frequency dependence of interaction propagators. Thus one can introduce an
additional scattering scalar potential given by Eqs. \ref{DeltaU}, \ref%
{DeltaH} substitute it into Eq. \ref{Inl}, and arrive at Eq. \ref{eq:InlGen}.

Generally $\beta_{eff}^{(1)}$ and $\beta_{eff}^{(2)}$ are phenomenological
parameters represented by thick wiggly lines in Fig.1a,d. At high electron
density, electrons are weakly interacting. In this limit the diagram in Fig.
1a is reduced to those shown in Fig.1b,c. These diagrams correspond to the
Hartree and Fock contributions respectively. This means that $%
\beta_{eff}^{(1,2)}$ can be calculated to first order in perturbation
theory with respect to the electron-electron interaction $V(\mathbf{r})$ 
\begin{eqnarray}
\beta_{eff}^{(1)}= V(0)-\frac{1}{2}\overline{V(k_{F}( \mathbf{\hat{n}- \hat{%
n^{\prime }}}))}  \nonumber \\
\beta_{eff}^{(2)}=\frac{1}{2}\overline{V(k_{F}(\mathbf{\ \hat{n}-\hat{
n^{\prime }}}))}  \label{eq:beta}
\end{eqnarray}
where $V(\mathbf{q})$ is a Fourier transform of a screened Coulomb
interaction. The bar denotes the average over the angle of the unit vectors $%
\mathbf{n}$, and $\mathbf{n}^{\prime}$. To verify this fact one has to show
that diagrams Fig.1e,f,g contains only combinations $\left( V(0)-\frac{1}{2}%
\overline{V(k_{F}( \mathbf{\hat{n}-\hat{n^{\prime }}}))}\right)^{2}$ and $%
\left(\frac{1}{2}\overline{V(k_{F}(\mathbf{\ \hat{n}-\hat{n^{\prime }}}))}%
\right)^{2}$.

The system of equations \ref{DeltaU},\ref{DeltaH},\ref{Inl},\ref{eq:beta} is
a generalization of that in Ref. \cite{SpivakZyuzin1} where only Hartree
term was taken into account. Eq. \ref{eq:beta} is typical for many effects
in mesoscopic conductors with interacting electrons (see for example \cite%
{AronovAltshuler,Aleiner}).  Usually, however, electron-electron
interaction effects give small corrections to the conductance of good conductors
with $G\gg e^{2}/\hbar$.  In our case, the magnitude of the effect is
proportional to $\beta_{eff}$.

To get Eq. \ref{eq:InlGen} one has to expand the expression for the
conductance in Eq. \ref{Inl} with respect to $\Delta u(\mathbf{r})$ and $%
\mathbf{h}(\mathbf{r})$.  To do so it is convenient to expand the potential 
\cite{ZyuzinSpivakNL} 
\begin{equation}
\Delta u(\mathbf{r})=\sum_{i}^{\infty }u_{i}n_{i}(\mathbf{r})
\end{equation}%
and the effective magnetic field 
\begin{equation}
h_{\alpha }(\mathbf{r})=\sum_{i}^{\infty }h_{\alpha ,i}n_{i}(\mathbf{r})
\end{equation}%
in a complete set of orthogonal eigenstates $n_{i}(\mathbf{r})$ of the
diffusion equation $(\int d\mathbf{r}n_{i}^{2}(\mathbf{r})=1)$: 
\begin{equation}
D\frac{\partial ^{2}}{\partial ^{2}\mathbf{r}}n_{i}(\mathbf{r})=E_{i}n_{i}(%
\mathbf{r})  \label{dif}
\end{equation}%
where $E_{i}$ are the eigenvalues of Eq.\ref{dif}, and $i$ labels the
eigenstates. We assume boundary conditions, which correspond to zero current
through a closed boundary, and $n_{i}(\mathbf{r})=0$ at the open
boundary. Generally speaking the electron density $\Delta n(\mathbf{r})$
and, consequently $\Delta u(\mathbf{r})$ contain all spatial harmonics, and
the problem is similar to the sensitivity of the sample conductance to a
change in the scattering potential $\Delta u(\mathbf{r})$ considered in \cite%
{LeeStone,AltshulerSpivak}. Thus we have 
\begin{equation}
F_{o}(\mathbf{H})=\frac{1}{2V}\sum_{i}\frac{dG(u_{i})}{du_{i}}(u_{i}(\mathbf{%
H})-u_{i}(\mathbf{-H}))+\frac{1}{2V}\sum_{i,\alpha }\frac{dG}{dh_{i,\alpha }}%
(h_{i,\alpha }(\mathbf{H})-h_{i,\alpha }(\mathbf{-H}))  \label{FbeforeAv}
\end{equation}%
In the absence of the spin-orbit scattering the second term in Eq.\ref%
{FbeforeAv} is zero while $dG/du_{i}$ is an even function of $\mathbf{H}$.
Thus in this case the effect originates from the odd in $\mathbf{H}$ part of 
$\Delta n(\mathbf{r},V,\mathbf{H})$. In the presence of the spin-orbit
scattering the second term is nonzero and there is another contribution to
the effect which comes from the odd-in-$\mathbf{H}$ part of $\mathbf{h}(%
\mathbf{r},V,\mathbf{H})$. Using Eq. \ref{FbeforeAv} and neglecting small
correlations between $\Delta G$, $\Delta u(\mathbf{r})$ and $\mathbf{h}(%
\mathbf{r})$ we get 
\begin{eqnarray}
\langle F_{o}^{2}\rangle &=&\frac{1}{4V^{2}}\sum_{i,j}\langle \frac{dG}{%
du_{i}}\frac{dG}{du_{j}}\rangle \langle (u_{i}(\mathbf{H})-u_{i}(-\mathbf{H}%
))(u_{j}(\mathbf{H})-u_{j}(-\mathbf{H}))\rangle +  \nonumber \\
&&\frac{1}{4V^{2}}\sum_{i,j,\alpha ,\beta }\langle \frac{dG}{dh_{i,\alpha }}%
\frac{dG}{dh_{j,\beta }}\rangle \langle (h_{i,\alpha }(\mathbf{H}%
)-h_{i,\alpha }(\mathbf{-H}))(h_{j,\beta }(\mathbf{H})-h_{j,\beta }(\mathbf{%
-H}))\rangle  \label{FafterAv}
\end{eqnarray}%
Eq. \ref{FafterAv} only contains correlation functions which can be
estimated in a single particle approximation (in zero order in $V(\mathbf{r}%
) $). This can be done in a standard way (see for example \cite%
{AronovAltshuler,LeeStone,SpivakZyuzin}) by calculating diagrams shown in
Fig.1e,f,g.  These diagrams contain ladder parts, shown in Fig.1i,j, which
depend on the electron spin indices. After summation over the spin indices, the
diagrams shown in Fig.1e,f,g only contain the blocks $P_{i}(\mathbf{r,r}_{1}%
\mathbf{,}\epsilon _{12}\mathbf{,H,}\omega _{s})$ $(i=1,2,3)$
described by the equations 
\begin{equation}
\begin{array}{c}
\left( -D\left( \mathbf{\nabla }+i\frac{e}{c}\mathbf{A}\right) ^{2}+i\omega
+i\omega _{s}sign\omega +\tau _{\varphi }^{-1}\right) P_{1}(\mathbf{%
r,r^{\prime },}\omega \mathbf{,H,}\omega _{s})+\frac{1}{2\tau _{so}}\left(
P_{1}-P_{2}\right) =\delta \left( \mathbf{r}-\mathbf{r}^{\prime }\right) \\ 
\left( -D\left( \mathbf{\nabla }+i\frac{e}{c}\mathbf{A}\right) ^{2}+i\omega
-i\omega _{s}sign\omega +\tau _{\varphi }^{-1}\right) P_{2}(\mathbf{%
r,r^{\prime },}\omega \mathbf{,H,}\omega _{s})+\frac{1}{2\tau _{so}}\left(
P_{2}-P_{1}\right) =0 \\ 
\left( -D\left( \mathbf{\nabla }+i\frac{e}{c}\mathbf{A}\right) ^{2}+i\omega +%
\frac{1}{\tau _{so}}+i\omega _{s}sign\omega +\tau _{\varphi }^{-1}\right)
P_{3}(\mathbf{r,r^{\prime },}\omega \mathbf{,H,}\omega _{s})=\delta \left( 
\mathbf{r}-\mathbf{r}^{\prime }\right)%
\end{array}%
\end{equation}%
Here $\tau _{so}$ is the spin-orbit mean free time. These equations account
for the dependence of $P_i$ on spin-orbit scattering and magnetic field. The magnetic
field enters through the vector potential $\mathbf{A}$ and the Zeeman
splitting $\omega _{s}=g\mu _{B}H$, where $\mu _{B}$ is the Bohr magneton.
The boundary condition for these diffusion poles at the insulating boundary is $\mathbf{n}\left( \mathbf{\nabla 
}+i\frac{e}{c}\mathbf{A}\right) P_{i}(\mathbf{r,r^{\prime },}\omega \mathbf{%
,H,}\omega _{s})=0$. In the case of ideal leads, when
the electron diffusion coefficient in the leads is infinite, the boundary
condition at the leads is $P_{i}(\mathbf{r,r^{\prime },}\omega \mathbf{,H,}%
\omega _{s})=0$. The quantities $P_{i}(\mathbf{r,r}_{1}\mathbf{,}\epsilon
_{12}\mathbf{,H,}\omega _{s})$ enter expressions for $I_{nl}$ only in the
combination 
\begin{equation}
\begin{array}{c}
\sum_{i=1,2}\left( P_{i}(\mathbf{r,r}_{1}\mathbf{,}\epsilon _{12}\mathbf{,}%
0,0)P_{i}(\mathbf{r}_{1}\mathbf{,r,}\epsilon _{21}\mathbf{,}0,0)-P_{i}(%
\mathbf{r,r}_{1}\mathbf{,}\epsilon _{12},2\mathbf{H,}\omega _{s})P_{i}(%
\mathbf{r}_{1}\mathbf{,r,}\epsilon _{21},2\mathbf{H,}\omega _{s})\right) +
\\ 
+P_{3}(\mathbf{r,r}_{1}\mathbf{,}\epsilon _{12}\mathbf{,}0,\omega _{s})P_{3}(%
\mathbf{r}_{1}\mathbf{,r,}\epsilon _{21}\mathbf{,}0,\omega _{s})-P_{3}(%
\mathbf{r,r}_{1}\mathbf{,}\epsilon _{12},3\mathbf{H,}0)P_{3}(\mathbf{r}_{1}%
\mathbf{,r,}\epsilon _{21},3\mathbf{H,}0)%
\end{array}%
\end{equation}%
It is convenient to choose the gauge $\mathbf{nA}=0$ at insulating boundary. In
this gauge, when calculating magnetic field dependence we can use standard
perturbation theory with respect to the magnetic field.

The sum over $i$ in Eq. \ref{FafterAv} converges quickly and, consequently,
the main contribution comes from the zero-harmonic with $i\sim 0$ (this fact is
related to the long range character of the correlation function of the part
of the electron densities which are proportional to $V$ \cite{SpivakZyuzin}
). Consequently the approximation where only this zero harmonic of the potential 
\begin{equation}
u_{0}(V,\mathbf{H})=\frac{\beta_{eff} }{v}\int_{v}\Delta n(\mathbf{r},V, 
\mathbf{H})d\mathbf{r}
\end{equation}
is taken into account gives a result valid in order of magnitude. In this
formula, $v$ is the volume of the sample, which reduces to $A$, the area of the sample, in the two-dimensional case.  In the case when $r_{s}\ll 1$ $%
\beta _{eff}^{\left( 1\right) }\gg \beta _{eff}^{\left( 2\right) }$ we also
can neglect the exchange field $\mathbf{h}( \mathbf{r},V,\mathbf{H})$ .

In the rest of the article we consider the case of a two-dimensional sample
which has a thickness much smaller that its lateral dimension, $L\gg L_{z}$. In
this case the results are different for cases of a perpendicular and a parallel
magnetic field.

\subsubsection{Effect in a Perpendicular Magnetic Field}

In the case where the magnetic field is oriented perpendicular to the film, spin-orbit
scattering can be neglected as long as $\tau _{so}\gg \tau $. In this case
the result depends on the relations between the sample size $L$, the
magnetic length $L_{H}$, the dephasing length\ $L_{\varphi }$, and the thermal coherence
length of normal metal $L_{T}$. It also depend on the nature of the leads to
the sample.

Let us start with the case of ideal leads when the electron diffusion
coefficient in the leads is infinite, $D_{L}=\infty $.

Then at small temperatures $L_{T},L_{\varphi }\gg L$ we get  
\begin{equation}
\langle F_{o}^{2}(\mathbf{H})\rangle =BH^{2}(\beta _{eff}^{(1)})^{2}\frac{%
e^{2}}{\hbar ^{2}\Gamma ^{4}A^{2}}\bigg(\frac{e^{2}}{h}\bigg)^{2}\bigg(\frac{%
L^{2}}{\Phi _{0}}\bigg)^{2}  \label{linHperp}
\end{equation}%
where $\Gamma =D/L^{2}$, $D=v_{F}l/2$ is the
diffusion coefficient inside a two-dimensional sample, $B$ is a
numerical factor of order one, and $A$ is the area of the sample.

The linear $H$-dependence of $I_{nl}$ given by Eq. \ref{linHperp} holds at
small magnetic fields when $HL^{2}\ll \Phi _{0}$, where $\Phi _{0}=\hbar e/c$
is the flux quanta. In the opposite limit $HL^{2}\gg \Phi _{0}$ the
function $F_{o}(\mathbf{H})$ exhibit random oscillations as a
function of $H$ with a characteristic period of order $\Delta H\sim \Phi
_{0}/L^{2}$. To verify this one can show that the correlation function $%
\langle F_{o}(H+\Delta H)F_{o}(H)\rangle $ decays at $\Delta H\sim \Phi
_{0}/L^{2}$. Qualitatively the dependence $F_{o}(\mathbf{H})$ in
the case $D_{L}=\infty $ is shown in Fig. 2a.

The temperature dependence of $F_{o}(\mathbf{H},T)$ is nontrivial because it
exhibit random oscillations as a function of $T$ which are shown
qualitatively in Fig. 2b. To verify this one has to calculate two
correlation functions 
\begin{equation}
\langle F_{o}^{2}(T)\rangle \sim \langle F_{o}(0)F_{o}(T)\rangle
\label{tempdep}
\end{equation}%
and notice that they have the same temperature dependencies. It also follows
from Eq. \ref{tempdep} that at $T>D/L^{2}$, the period of oscillations is on the order of the temperature itself. Let us turn now to the case when the leads have
the same diffusion coefficient as the sample, $D_{L}=D$. In this case $H$ and 
$T$ dependencies of $F(H,T)$ are qualitatively shown in Figs. 2c,d. The main
difference with the case $D_{L}=0$ is that in this case $F_{o}(T,H)$
exhibits random oscillations as functions of $H$ and $T$ even in the
case $L_{H}\gg L$ and $L_{T}\gg T$. These oscillations are related to the
existence of diffusive electron trajectories which leave the sample, travel
in the leads and then come back. Their contribution to the total current is
small, but their sensitivity to changes of $H$ and $T$ are so big that the
derivatives $\partial F_{o}/\partial T\rightarrow \infty $ and $
\partial F_{o}/\partial H\rightarrow \infty $ at zero temperature and
magnetic field $H\rightarrow 0$. Qualitatively the $T$ and $H$ dependencies
of $F_{o}(T,H)$ are similar of those of the linear conductance \cite
{LeeStone,ZyuzinSpivakH,CobdenSpivakZyuzin}.

\subsubsection{Effect in a Parallel Magnetic Field}

When deriving Eq. \ref{linHperp} we neglected the effect of Zeemann
splitting of the electron spectrum because it gives a small contribution to
the effect. However, when a thin enough sample is oriented parallel to a
magnetic field $\mathbf{H}_{\Vert}$, the orbital contribution of the
magnetic filed is absent, and Zeemann splitting causes the dominant
contribution to $I_{nl}$. However, in the absence of spin-orbit scattering, $%
\delta n(\mathbf{H}_{\Vert})$ , and consequently $I_{nl}$ are even functions
of $\mathbf{H}_{\Vert}$. Thus in this case the effect is determined by the
spin-orbit scattering rate $1/\tau _{so}$. In the presence of spin-orbit
scattering the dashed lines in Fig.1 correspond to the following expression 
\begin{equation}
\frac{1}{2\pi \nu \tau }\delta _{\alpha \beta }\delta _{\delta \gamma }+ 
\frac{1}{8\pi \nu \tau _{so}}\mathbf{\sigma }_{\alpha \beta }\mathbf{\sigma }
_{\delta \gamma }  \label{Dashedline}
\end{equation}

In the case of a weak magnetic field, $\mathbf{\mu _{B}\cdot H_{\Vert }}\ll
1/\tau _{so},\Gamma$, and at $T=0$, we get an expression for the odd-in-$
\mathbf{H}_{\Vert }$ part of the zero mode of the density fluctuation
\begin{equation}
\langle F_{o}^{2}(\mathbf{H}_{\Vert })\rangle =C\left( \frac{e^{2}}{h}
\right) ^{2}\frac{e^{2}}{\hbar ^{2}A^{2}\Gamma ^{2}}\left( \beta
_{eff}^{(1)}\right) ^{2}\frac{1}{\tau _{so}}\frac{1}{\Gamma \left( \Gamma +
\frac{1}{\tau _{so}}\right) ^{4}}\left[ 2\Gamma ^{2}+2\Gamma \frac{1}{\tau
_{so}}+\frac{1}{\tau _{so}^{2}}\right] \left( \frac{\mu _{B}\mathbf{H}
_{\Vert }}{\Gamma }\right) ^{2}
\end{equation}
where $C\sim 1$. In the case of weak spin-orbit scattering $\Gamma \gg \tau
_{so}^{-1}$ we have $\langle F_{o}^{2}\rangle =\Gamma ^{-7}$. It
is interesting that the amplitude of the effect is proportional to the
amplitude of the spin-orbit scattering rather than it's rate. This
expression holds as at $\mu _{B}H_{\Vert }<\Gamma $. In the opposite limit
we have $\langle F_{o}^{2}\rangle =\Gamma ^{-5}$ which is
independent of $\mathbf{H}_{\Vert }$. In this region $F_{o}(\mathbf{H}%
_{\Vert })$ exhibit random sample specific oscillations as a function
of $H_{\Vert }$ with a characteristic period of order $\Gamma /\mu _{B}$. To
get this result one has to calculate the correlation function $\langle 
\left[ F_{o}(\mathbf{H}_{\Vert })-F_{o}(\mathbf{H_{\Vert }+\Delta H_{\Vert }}
)\right] ^{2}\rangle $ and to see that it decays at $\mu H_{\Vert
}>\Gamma $.

In the limit of a strong spin-orbit scattering $\Gamma ,\mu _{B}\mathbf{H}
_{\Vert }\ll 1/\tau _{so}$ we have $\langle F_{o}^{2}\rangle \sim
1/\tau _{so}$ , and the amplitude of the effect which decreases as $
\tau _{so}$ increases. In the limit of strong magnetic field $\mu _{B}
\mathbf{H}_{\Vert }\gg 1$ we have $\langle F_{o}^{2}\rangle =\tau
_{so}^{2}/\Gamma ^{5}$. In this case the period of the oscillations
of $F_{o}(\mathbf{H}_{\Vert })$ as a function of $\mathbf{H}
_{\Vert }$ is of order $1/\tau _{so}$.

\section{Conclusion}

We have shown that odd-in-$\mathbf{H}$ and quadratic-in-$V$ part of the
current is proportional to the electron-electron interaction constant. Thus
detailed measurement of this effect should yield information about the
strength of the electron interaction. On the other hand, in mesoscopic
samples the amplitude of the current given by Eq. 3 is proportional to a random sample-specific sign. Thus it is unclear at the moment
whether it is feasible to extract the sign of the electron interaction from
measurements of the nonlinear current Eq. \ref{eq:InlGen}.

At some level the effect considered above is similar to the effect of the
interactional corrections to average conductivity of disordered metals \cite
{AronovAltshuler}. These corrections originate from a correlation between
random electron diffusive trajectories and the Friedel oscillations of the
electron density in the presence of random potential. The difference is the
following. In the case of Ref. \cite{AronovAltshuler} electrons scatter on
equilibrium Friedel oscillations of the electron density, which are even in $
\mathbf{H}$. At $G\gg e^{2}/h$ this leads to non-analytic but small and even
in $\mathbf{H}$ corrections to the Drude conductance. The origin of our
effect is the electron scattering on non-equilibrium fluctuations of the
electron density. In this case the electron interaction determines the
magnitude of the effect. The effect considered above is also different from
classical effects \cite{Photovoltaic,ivchenko} the amplitude of which is
proportional to the inelastic relaxation rate and which does not exhibit
oscillations as a function of temperature, magnetic field and the chemical
potential.

The magnitude of the effect discussed above decays as the temperature
increases.
 If the crystalline
structure of the material is non-centro-symmetric, at high enough
temperatures, the $T$-dependence of $F_{o}(\mathbf{H},T)$ is determined by
the "classical" effects considered in Ref. \cite
{Photovoltaic,BelinicherIvchenko,ivchenko}. However, if the structure of the
pure crystal is centro-symmetric, then the only source of the effect at high temperatures is the
non-centro-symmetric distribution of the scattering potential 
$u(\mathbf{r} )$. 
In this case the effect is of a classical nature.
 Namely, one should consider the
classical motion of interacting electrons in the presence of a frozen random
potential, similar to what has been done for average quantities in \cite
{Poliakov}. This problem however, is beyond the scope of this article.

Finally we would like to discuss a difference between our approach and the
approach in Ref. \cite{Buttiker}. At $T=0$ and in the absence of the
electron interaction the Landauer formula in a combination with the random
matrix theory has been a useful tool for describing the linear conductance
of mesoscopic samples. In this approximation it can be derived from the Kubo
formula \cite{LeeFisher}. However, even in this case the derivation can be
carried out only in the case of ideal leads, $D_{L}=\infty$, when incident and
transmitted waves through the sample are well defined. At finite
temperature, and in the absence of inelastic processes $L\ll L_{\phi}$
equation \ref{landauer} still can be applied. However, to describe the
temperature dependence of the conductance $G(T)$ one has to know delicate
properties of the energy dependencies of the matrix elements $%
T_{ij}(\epsilon)$. At even higher temperatures, $L\gg \sqrt{D\tau_{\epsilon}}
$, when inelastic scattering processes are significant, Eq. \ref{landauer}
cannot be applied. One of the reasons for this is that the electron channels
(and even their number) in Eq. \ref{landauer} are not well defined in this
case. For example, the Landauer formula cannot reproduce the well-known
Bloch $T^{5}$ and $T^{2}$ temperature dependencies of resistivity of bulk
metals associated with electron-electron and electron-phonon scattering \cite
{AbrikosovMetals}. It also can not reproduce electron interaction
corrections to the conductance considered in \cite{AronovAltshuler}.

Sometimes the Landauer approach gives correct results for linear conductance
of samples even in the case when the leads are not ideal and $D_{L} <
\infty$. Consider for example a constriction between two semi-infinite 3D
metals and assume that the diffusion coefficient is independent of the
coordinates. The Landauer formula still gives a correct result for the
conductance of the sample in a single particle approximation because the conductance is
determined by the part of the sample near the constriction. On the other
hand, the magnetic field $G(\mathbf{H}+\delta \mathbf{H})-G(\mathbf{H})$ and
the temperature $G(T+\delta T)-G(T)$ dependencies of the conductance in this
case are determined by the interference of diffusive paths travelling on
distances of order of the magnetic length $L_{\delta H}$ and the coherence
length $L_{\delta T}$ of the normal metal. At small $\Delta T$ and $\delta 
\mathbf{H} $ these lengths are much bigger than the constriction size, and
they diverge as $\delta H, \delta T \rightarrow 0$. As a result, in the case
of non-ideal leads, $D_{L}<\infty$, and in the absence of inelastic phase
breaking processes, the periods of oscillations of $G(H,T)$ as a function of 
$\delta T$ and $\delta \mathbf{H}$ decrease at small $H$ and $T$, and the
derivatives $d G/dH$ and $d G/ dT$ diverge at $T\rightarrow 0$ and $
H\rightarrow 0$ \cite{SpivakZyuzin}. To cut off these divergencies one has
to take into account inelastic electron scattering processes. Such effects
are beyond the Landauer formula.

The situation with non-linear parts of the I-V characteristics is more
complicated. In the single particle approximation the I-V characteristics
can be expressed in terms of the $\epsilon$-dependence of the matrix
elements $T_{ij}(\epsilon)$, where $\epsilon$ is the electron energy. This
procedure gives a result equivalent to that obtained in \cite%
{LarkinKhmelnitskii} using Keldysh diagram technique. In this approximation,
however, $F_{o}(H)=0$. Generally speaking in the presence of the electron
interaction the Landauer approach can not be justified even in the case of
ideal leads, and even at $T=0$. One of the reasons is that in the presence
of electron-electron and electron-phonon inelastic processes allowed at $
V\neq 0$, the electron channels are not well defined. In other words the
voltage plays the role similar to the temperature, and at $V\neq
0$ there exist effects which are similar to $T$-dependent interactional
corrections to conductivity \cite{AronovAltshuler}. On the other hand, at
small voltages these processes lead to the value of $I_{nl}$ proportional to
a power of V greater than or equal to three. Then the question arises
whether the Landauer scheme can be modified to describe the quadratic-in-$V$
current of Eq. \ref{eq:InlGen}. The authors of Ref. \cite{Buttiker}
introduced a concept of non-equilibrium capacitance which relates $V$ and
the total charge induced in the sample. On a phenomenological level this
approach is similar to those presented in Ref. \cite{SpivakZyuzin1} and in
this article. We would however like to mention, the differences. The
approach of Ref. \cite{Buttiker} corresponds to accounting for only the zero
harmonics of $\Delta u(\mathbf{r})$ in Eq. \ref{Inl}. Though this
approximation is not exact, in diffusive samples and at $T=0$ it gives the
correct order of magnitude of the effect. In ballistic quantum dots the
mistake is much bigger. More importantly the approach of Refs. \cite
{Buttiker} is restricted to the Hartree approximation and can not account
for the exchange interaction. We would like to mention that before averaging
over realizations of the scattering potential, the exchange terms in Eqs. \ref
{DeltaU} and \ref{DeltaH} can change even the sign of the effect.

In conclusion we would like to mention that there must also exist currents
through the sample which are proportional to $H(\nabla T)^{2}$ and $HV
\nabla T$ .

This work was supported by the National Science Foundation under Contracts
No. DMR-0228104, and by the Russian Fund for Fundamental Research 05-02-17816a.
We thank A. Andreev, P. Brouwer, D. Cobden, C. M. Marcus, D. M. Zumbuhl, and
H. Bouchiat for useful discussions.

\newpage

\begin{figure}[ptb]
\includegraphics[scale=0.7,bb=0 0 1099 792]{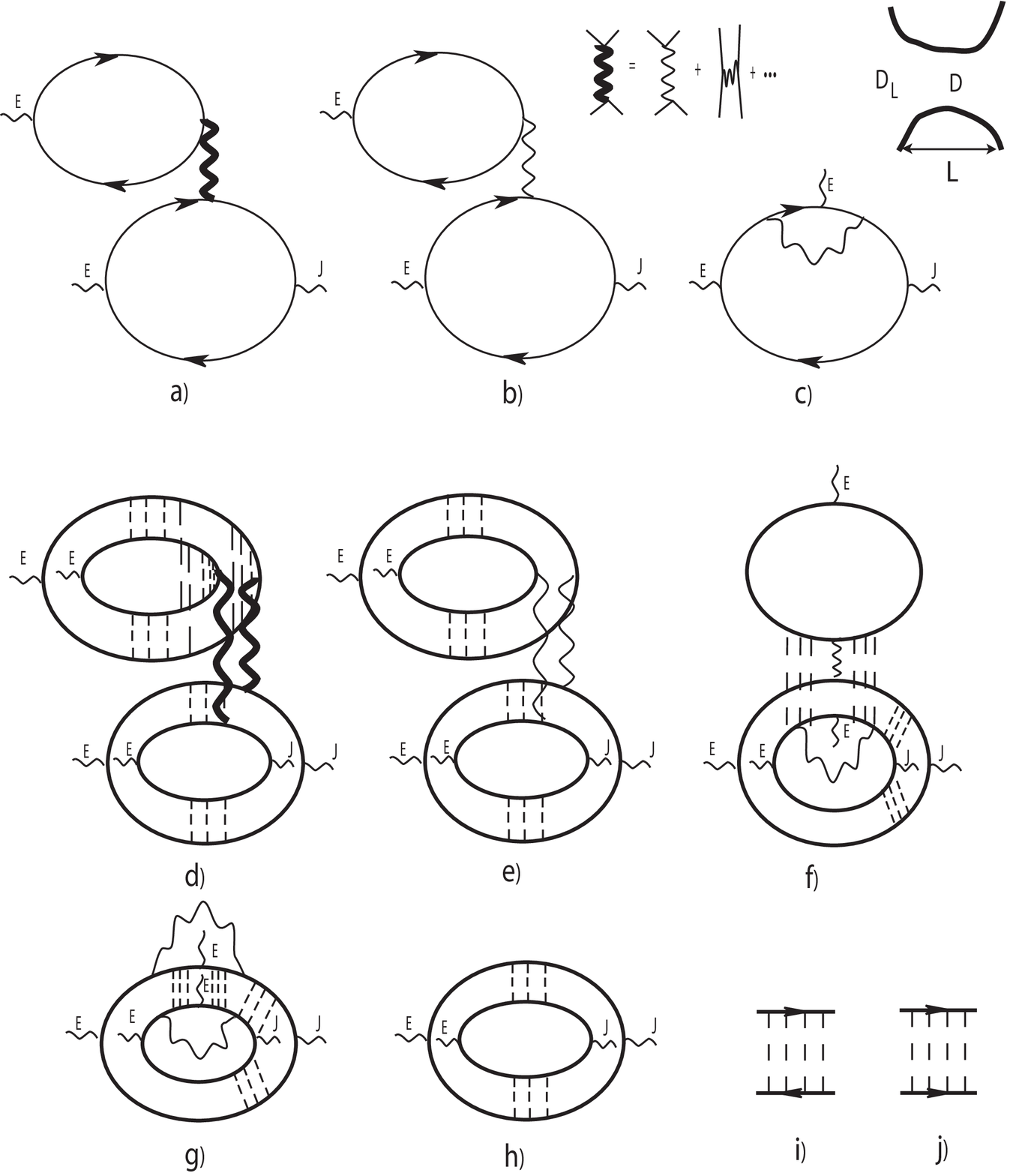}
\par
\label{fig:plot}
\caption{Diagrams describing $I_{nl}$, and $\langle I_{nl}^{2}\rangle$. Thin
thin solid lines correspond to the electron green functions, thick solid
lines correspond to electron Green functions averaged over random
realizations of the scattering potential, dashed lines correspond to the
correlation function of the scattering potential given by Eq. \protect\ref
{Dashedline} . The fat wiggly line is the effective electron interaction
interaction, thick wiggly lines correspond to the elecron-electron
interaction $V(\mathbf{r})$. Diagrams b,c corresponds to Hartree (\textbf{b}
) and Fock (\textbf{c}) terms of the perturbation theory with respect to $V(
\mathbf{r})$.}
\label{fig:main}
\end{figure}

\newpage

\begin{figure}[ptb]
\includegraphics[scale=0.5,bb=0 0 544 735]{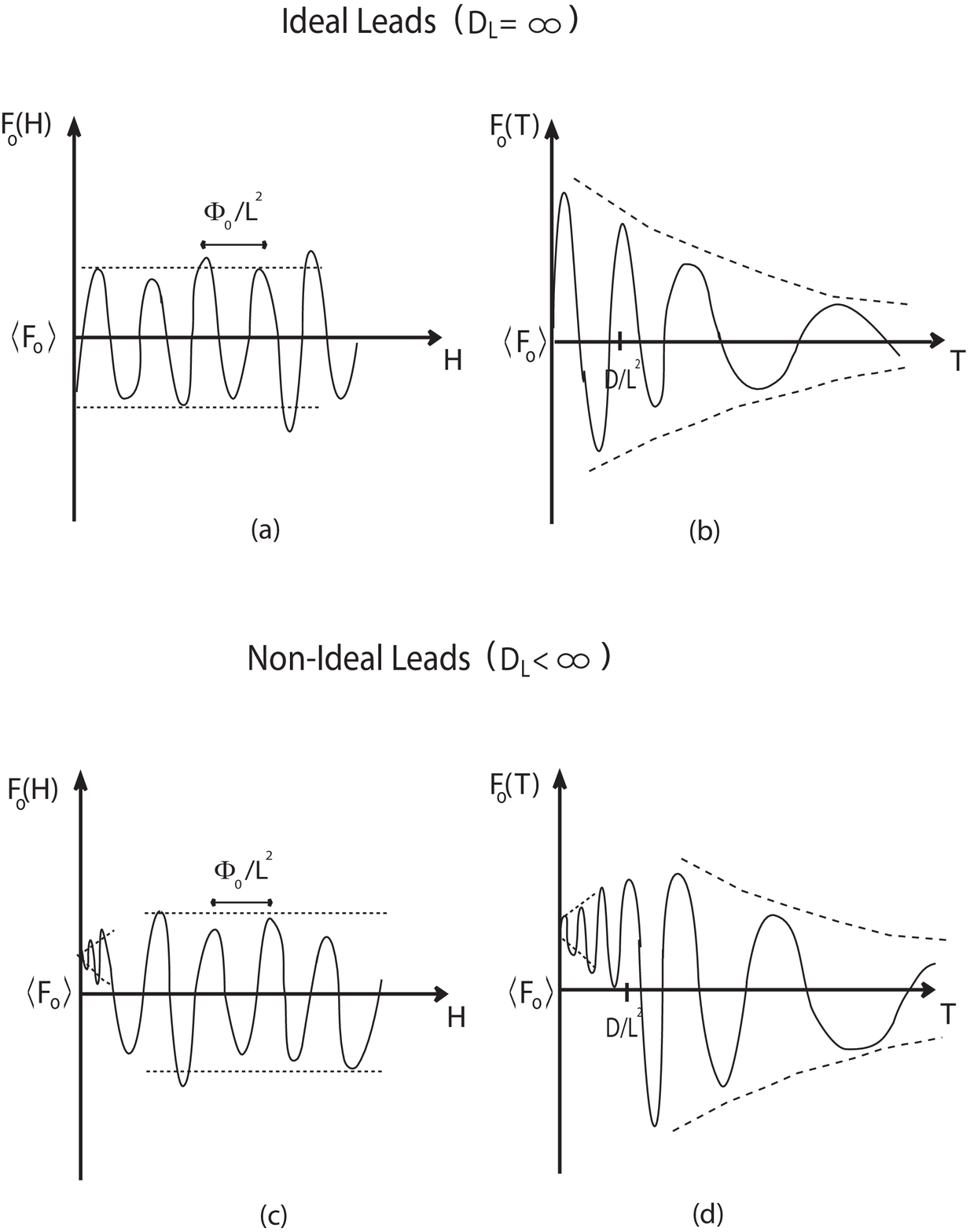}
\par
\label{fig:plot}
\caption{Qualitative pictures of $H$- and $T$ dependencies of $F_{0}(H,T)$.
Figs. a) and b) correspond to the case of ideal leads ($D_{L}=\infty$),
while Figs. c and d correspond to the case of diffusive leads $D_{L}<\infty$
. }
\label{fig:main}
\end{figure}

\end{document}